# Superconductivity in the crystallogenide LaFeSiO$_{1-\delta}$ with squeezed FeSi layers

M. F. Hansen,[1] J.-B. Vaney,[2] C. Lepoittevin,[1] F. Bernardini,[3] E. Gaudin,[2] V. Nassif,[1, 4] M.-A. Méasson,[1] A. Sulpice,[1] H. Mayaffre,[5] M.-H. Julien,[5] S. Tencé,[2] A. Cano,[1] and P. Toulemonde[1, *]

[1]*CNRS, Université Grenoble Alpes, Institut Néel, 38000, Grenoble, France*
[2]*CNRS, Université Bordeaux, ICMCB, UPR 9048, F-33600 Pessac, France*
[3]*Dipartimento di Fisica, Università di Cagliari, IT-09042 Monserrato, Italy*
[4]*Institut Laue -Langevin, 71 Avenue des Martyrs, 38000 Grenoble, cedex 9 France*
[5]*CNRS, LNCMI, Université Grenoble Alpes, INSA-T, UPS, EMFL, Grenoble, France*
(Dated: June 23, 2022)

Pnictogens and chalcogens are both viable anions for promoting Fe-based superconductivity and intense research activity in the related families has established systematic correlation between the Fe-anion height and the superconducting critical temperature $T_c$, with an optimum Fe-anion height of $\sim 1.38$ Å. Here, we report the discovery of superconductivity in a novel compound LaFeSiO$_{1-\delta}$ that incorporates a crystallogen element, Si, and challenges the above picture: considering the strongly squeezed Fe-Si height of 0.94 Å, the superconducting transition at $T_c = 10$ K is unusually high. In the normal state, the resistivity displays non-Fermi-liquid behavior while NMR experiments evidence weak antiferromagnetic fluctuations. According to first-principles calculations, the Fermi surface of this material is dominated by hole pockets without nesting properties, which explains the strongly suppressed tendency towards magnetic order and suggests that the emergence of superconductivity materializes in a distinct set-up, as compared to the standard $s_{\pm}$- and $d$-wave electron-pocket-based situations. These properties and its simple-to-implement synthesis make LaFeSiO$_{1-\delta}$ a particularly promising platform to study the interplay between structure, electron correlations and superconductivity.

## INTRODUCTION

Iron-based superconductors (IBSC) are presently a well established class of unconventional superconductors, spanning multiple structural families [1]. At their core, IBSCs consist of a square planar lattice of Fe atoms, tetrahedrally coordinated by pnictogen or chalcogen elements $X$ (typically $X$ = As or Se) placed above and below the Fe plane. Different spacers can be intercalated between this central structural unit, thereby forming the different IBSC families. It turns out that the superconducting critical temperature ($T_c$) can be correlated with the $X$ anion height from the Fe plane $h_{Fe-X}$ (shown in Fig. 2b), with the maximum T$_c \simeq 56$ K corresponding to $h_{Fe-X} \sim 1.38$ Å [2, 3]. At the optimal height, the FeX$_4$ tetrahedra becomes regular with the $X$-Fe-$X$ angle $\alpha$ = 109.47°. The IBSC class has recently been extended to other layered materials where pnictogen/chalcogen atoms are replaced by Ge in YFe$_2$Ge$_2$ and by Si in LaFeSiH and LaFeSiF$_x$ [4–6]. This is very interesting because it further establishes the prospect of finding more IBSCs. Simultaneously, it is also surprising since the use of crystallogens —i.e. group 14 elements— has been discussed as detrimental to superconductivity since a ferromagnetic ground state should become favoured as opposed to the anti-ferromagnetic ground state which is normally associated with the parent compounds of IBSCs [7].

In this work, we report a further extension of the iron-crystallogen superconductors with the synthesis of the novel compound LaFeSiO$_{1-\delta}$ . This compound represents a new intriguing member of the so-called 1111 family not only for evading ferromagnetism but also, and perhaps more importantly, because of its exceptional crystal structure (Fig. 2a) where the Fe-Si height drops to 0.94(1) Å. This parameter is far away from what is considered the optimal geometry for superconductivity in the IBSCs, and indeed produces drastic changes in the electronic properties as we show below. Yet, superconductivity is observed below onset $T_c = 10$ K with a small $\delta = 10$ % oxygen deficit in the compound. This finding thus challenges the current notion that crystallogens should be avoided when searching for new IBSCs, and provides a qualitatively new platform to further scrutinize the link between the crystal structure and the electronic properties in Fe-based superconductors.

## RESULTS

### Synthesis and Crystallographic structure

Polycrystalline LaFeSiO$_{1-\delta}$ samples were synthesized from the non-superconducting, weak Pauli paramagnet LaFeSi precursor [8] (see details in Methods). Fig. 1 shows the energy dispersive X-ray spectroscopy (EDX) spectrum of one crystallite measured after oxygenation in a transmission electron microscope (TEM). The elemental composition deduced from the spectrum confirms the atomic ratio 1:1:1:1 for the elements La, Fe and Si as in the precursor. In addition, we observe an intense peak at 0.525 keV. This peak corresponds to the K$_{\alpha_1}$ electronic transition of oxygen, which shows that oxygen is present. The inset in Fig. 1 shows the $hk0$-plane of the 3D electron diffraction patterns obtained from the crystallite. This cut reveals that the tetragonal *P4/nmm* space-group symmetry of the precursor is preserved after oxygenation. Furthermore, the analysis of the full dataset reveals unambiguously electron density corresponding to oxygen, occupying the 2*b* Wyckoff position located at the center of the La$_4$ tetrahedron (see Fig. 6 in Supplementary Information (S.I.)). This shows that the oxygen which was detected by EDX is in fact present in the crystal structure and is not only a surface contamination.

The refined structural parameters at 300 K from neutron powder diffraction (NPD) are shown in Table I (corresponding to the Rietveld fit shown Fig. 2b). The refinement confirms the presence of oxygen in the La$_4$ tetrahedron with an occupancy of 0.90(2). Investigating several samples has shown similar values of the unit cell parameters, indicating that this occupancy is consistently reached.

When comparing to the precursor [8, 9], we observe a strongly anisotropic expansion of the unit cell, due to the oxygen inserted in the iso-symmetrical (*P4/nmm*) structure of LaFeSi. Specifically, the lattice parameters of LaFeSiO$_{1-\delta}$ are *a* = 4.1085(4) Å and *c* = 8.132(2) Å, resulting in a change of the *c*/*a* ratio from LaFeSi to LaFeSiO$_{1-\delta}$, from 1.74 to 1.98.

The refined atomic positions, also shown in Table I, reveal that the *z* parameter of Si is low, leading to an anion height $h_{Fe-Si}$ = 0.94(1) Å. This is considerably lower than in LaFeAsO where $h_{Fe-As}$ = 1.32(1) Å [10, 11]. In LaFeSiO$_{1-\delta}$, however, the Fe-containing block is chemically different from the arsenides so it may be better compared to LaFeSiH for which $h_{Fe-Si}$ = 1.20(1) Å [4]. In any case, the trend is similar to what is seen for the arsenides where the Fe-As layer is more compressed in LaFeAsO than the substituted LaFeAsO$_{1-x}$H$_x$ [10]. However, $h_{Fe-Si}$ in LaFeSiO$_{1-\delta}$ is far from the geometries recorded for other IBSCs. Considering the correlation between $h_{Fe-Si}$ and $T_c$ currently proposed in the literature for iron pnictides or -chalcogenides, this geometry should be detrimental to superconductivity [12].

It is also interesting to consider the resulting angle $\alpha$(Si-Fe-Si) of the FeSi$_4$ tetrahedral unit as this is often used as a measure of the tetrahedral geometry. The $T_c$ is normally optimised around the regular tetrahedral value namely $\alpha$ = 109.47° [11, 13–15]. In LaFeSiO$_{1-\delta}$ the $\alpha$ angle is found to be $\alpha$ = 130.9 (8)°, resulting from the compression of the Fe-Si layer along the *c*-axis upon insertion of oxygen. This is again far away from the geometry where superconductivity is optimized for arsenides and it is also away from values found in the fringe case LaFePO where $T_c$ is below 6 K with an angle of $\alpha$ = 119.2° [16–18].

The crystal structure was also measured at low temperatures (2 K and 100 K) using NPD. The unit cell is contracted at low temperatures and unit cell parameters reach a = 4.1019(6) Å and c = 8.090(2) Å at 2 K (see inset of Fig. 2b). In these measurements, no signature of neither structural distortion nor long-range magnetic order was detected.

## Superconducting properties

Fig. 3a shows the electrical resistance as a function of temperature as measured on a small grain of LaFeSiO$_{1-\delta}$. The residual resistivity ratio of this grain is around 15 (Fig. 3b), much better than for large cold pressed samples (~ 2). When measuring these large samples we observe a drop in the resistance at low temperatures, which is partial, likely due to insulating grain boundary effects (see Fig. 9 in S.I.). However, for the small grain, the drop is complete as we can see in Fig. 3b. This evidences superconductivity in LaFeSiO$_{1-\delta}$ with onset $T_c \simeq$ 10 K.

In Fig. 3c we show the field dependence of $T_c$ up to 7 T determined for a large cold pressed polycrystalline sample. By performing a linear fit and using the Werthamer-Helfand Hohenberg (WHH) formula, we roughly estimate the upper critical field $H_{c2}$(0 K) to be ~ 17 T. Since the $T_c$ is determined from the onset and not from zero resistivity, such value should be considered as an upper bound for the true thermodynamic $H_{c2}$.

Fig. 3d shows the magnetization difference $\Delta M$ between 2 K and 15 K, i.e. below and just above $T_c$, as a function of field. The typical hysteresis loop of a type-II superconductor is clearly observed. There is also a significant ferromagnetic contribution that saturates around 3 T. Nevertheless, the change in magnetization between 2 K and 15 K is dominated by the superconducting phase, whereas the ferromagnetic contribution, attributed to the secondary phase La(Fe$_{1-x}$Si$_x$)$_{13}$ (see S.I. for a detailed discussion), changes very little at low $T$ given its very high Curie temperature $T_{Curie}$ > 200 K [19].

Evidence of diamagnetism is further provided by the negative sign of the zero-field cooled susceptibility, as measured by the slope of $\Delta M$ ($H$) (Fig. 3e). The volumetric susceptibility calculated from the slope is $\chi_V$ = −0.15 which corresponds to 300 times the diamagnetic susceptibility of pyrolytic carbon, the strongest non-superconducting diamagnetic substance known in the literature [20]. Therefore, the diamagnetic signal observed in these measurements again evidences superconductivity in LaFeSiO$_{1-\delta}$.

We note that the measured volume susceptibility is a linear combination of contributions from the superconducting LaFeSiO$_{1-\delta}$ phase and the ferromagnetic background, the two having opposite signs. Therefore, the apparent susceptibility yields just a lower bound for the estimate of the superconducting volume fraction, which is 15%. Such moderate superconducting volume fraction, as well as the relatively broad transition observed in the resistivity, may also be linked to chemical inhomogeneity on the oxygen site arising from the sluggish nature of the oxygenation process. Hence, despite the refined oxygen deficit of the sample being $\delta$ = 0.1, the actual oxygen content behind the observed superconductivity at ~10 K may be different and distributed over some small $\delta$ range. However, the slight optimization of $T_c$ in this oxysilicide, reachable by tuning $\delta$, is out of the scope of the present study. In any case, the superconducting volume fraction measured in our samples is definitely larger than the amount of secondary phases so that it can safely be attributed to LaFeSiO$_{1-\delta}$ (for more details see S.I.).

Around 30 Oe the magnetization deviates from the linear behaviour observed at low field (Fig. 3e). This gives us an estimate of the lower critical field $H_{c1}$ at 2 K. The magnetization as a function of temperature was also measured, and is shown in Fig. 3f. We observe the Meissner effect as well as a large shielding around 10 K (see zoom in Fig. 8 in S.I.) despite the magnetic background from parasitic phases contributing as a linear slope in the magnetization.

## Normal-state properties

In the normal state, the resistivity varies as $T^{\alpha}$ as found in other Fe-based superconductors [21–23]. While enlarging the fitting range tends to decrease $\alpha$ and to degrade the fit quality, a good fit to $T^{1.4}$ is obtained up to 80 K as shown in Fig. 3a. Finding $\alpha' = 2$ is typical of non Fermi-liquid behavior. Considering the established correlation between the resistivity exponent $\alpha$ and the strength of spin fluctuations [24–26], the value $\alpha \simeq 1.4$ suggests that charge carriers in LaFeSiO$_{1-\delta}$ are scattered off spin fluctuations of similar strength as moderately overdoped Fe-based pnictides, tetragonal FeSe$_{1-x}$S$_x$ [27] or YFe$_2$Ge$_2$ [28]. As we now explain, the presence of spin fluctuations is supported by our $^{29}$Si nuclear magnetic resonance (NMR) results.

First, we observe that the Knight shift $K$ decreases from room $T$ down to low $T$ (Fig. 4a). As in most Fe-based superconductors of various doping levels, this behavior reflects the decrease of the static, uniform spin susceptibility $\chi_{\text{spin}}(q = 0)$ upon cooling (see for example refs. [25, 26, 29–32]). Visibly, the Fe $d$ electrons produce a transferred hyperfine field at Si sites, just as they do at As/Se sites in iron pnictides/chalcogenides. $^{29}$Si NMR thus promises to be a sensitive probe of the electronic properties in this new family of Fe-based superconductors.

Here, in LaFeSiO$_{1-\delta}$, we find that the spin-lattice relaxation rate $1/T_1$ divided by temperature $T$ increases at low $T$ (Fig. 4c), which signifies that the low-energy ($\sim \mu$eV) spin fluctuations strengthen upon cooling. The observed $\sim 50\%$ enhancement of $1/(T_1 T)$ resembles data in the middle of the overdoped regime of 1111 or 122 families of iron-based superconductors, for which spin fluctuations are relatively weak [25, 26, 29–38]. This observation is consistent with the above described resistivity exponent but one should not conclude from this that LaFeSiO$_{1-\delta}$ has the same doping or the same Fermi surface as moderately overdoped 122 pnictides: for instance, a similarly mild enhancement of $1/(T_1 T)$ is also found in nonsuperconducting Fe$_{1.03}$Se [39], in tetragonal FeSe$_{1-x}$S$_x$ [31], in LiFeAs [26] and in LiFeP [40]. On the other hand, Fe pnictides with -or close to- spin ordering [25, 26, 29–38], or even FeSe that does not order [39], show much larger enhancement of $1/(T_1 T)$ at low $T$. The relatively weak, albeit tangible, spin fluctuations imply that LaFeSiO$_{1-\delta}$ does not lie in the immediate vicinity of a magnetic instability. A difference with FeSe$_{1-x}$S$_x$ and most 1111 or 122 pnictides (a notable exception being BaFe$_2$(As$_{1-x}$P$_x$)$_2$ [24]) is the absence of any discernible activated contribution to $1/(T_1 T)$ at high temperatures (typically between 300 K and 100 K), usually attributed to small-momentum fluctuations (so-called intraband transitions [29]). This thus suggests distinctive Fermi surface topology in LaFeSiO$_{1-\delta}$.

The dominant wave vector $\mathbf{q}$ of the fluctuations cannot be determined from the present experiment, so it is not necessarily $(0, \pi)$ in principle. Nevertheless, that the value of the ratio $\alpha_{\text{Korr}} = \hbar \gamma_e^2/(4\pi k_B \gamma_n^2)(1/(T_1 T K^2))$ (where $\gamma_e$ and $\gamma_n$ are the electron and nuclear gyromagnetic ratio, respectively) largely exceeds 1 and grows upon cooling (Fig. 4f) is an indication that the fluctuations are predominantly of antiferromagnetic nature [41], *i.e.* with $\mathbf{q} \neq 0$ (note that in this estimate we have implicitly assumed that the orbital contribution to $K$ is small compared to the spin contribution and that the hyperfine field at Si sites is relatively isotropic).

The NMR data also provide evidence of spatial heterogeneity, as observed in several Fe-based materials [33–38]: upon cooling below $\sim$100 K, the moderate increase of the line width (Fig. 4b) indicates that the distribution of Knight shift values broadens. The concomitant deviation from 1 of the stretching exponent $\beta$ (Fig. 4d and Methods) shows that a distribution of $T_1$ values develops alongside with the growth of spin fluctuations. The distributed $K$ and $T_1$ likely stem from spatial variations of the electronic spin polarization around defects [42, 43].

Finally, we notice that $1/(T_1 T)$ no longer increases below 20 K and even drops somewhat below 10 K, that is, below a temperature close to the zero-field $T_c$ (Fig. 4c). This is surprising since the magnetic field of 15 T used in the NMR experiment should be close to the superconducting upper critical field $H_{c2}$ (see above) and thus we would expect to see essentially no sign of superconductivity down to our lowest temperature of 1.7 K. That the stretching exponent $\beta$ concomitantly reverts its $T$ dependence (Fig. 4d) suggests that both the spectral weight and the inhomogeneity of low-energy spin fluctuations are reduced below 10 K. This behavior is unlikely to arise from inhomogeneous superconductivity in the sample or from freezing of spin fluctuations at the NMR timescale [33] as both mechanisms should not lessen the inhomogeneity. More work is however required to understand this interesting pseudogap-like behavior that parallels earlier observations in LiFeP [40] as well as in Co and F-doped LaFeAsO [37, 38, 44]) and FeSe$_{1-x}$S$_x$ [31, 45].

## Electronic structure

Fig. 5 shows the calculated orbital-resolved density of states (DOS) and the band structure of LaFeSiO. Similarly to the reference LaFeAsO compound [46], there is a group of 12 bands between $-5.5$ eV and 2.5 eV relative to the Fermi energy $E_F$ that come from O-$2p$, Si-$2p$ and Fe-$3d$ states, with the La states contributing at higher energy. The Fe-$3d$ derived bands, in particular, appear between $-2.5$ eV and 2 eV and dominate the DOS at the Fermi level and thereby the metallic character of the system. However, the distinct crystal structure of LaFeSiO has a fundamental impact on the low-energy electronic features of this new material. While the Fermi surface preserves the two hole cylinders around the Brillouin zone center (i.e. around the $\Gamma$-Z direction), the extra band that crosses the Fermi level and gives rise to the third 3D hole pocket in LaFeAsO is pushed upwards at higher energy. In this way, the hole doping introduced by the As → Si substitution is absorbed in a non-rigid-band-shift fashion and results in tiny electron pockets at the zone edge (M-A line). When it comes to superconductivity, however, the band that mainly absorbs this doping remains passive in the standard picture (see e.g. [47]). Moreover, one of the electron pockets around the M-A line looses its Fe-$3d_{x^2-y^2}$ content in favor of a Si-$2p$ character due to the hybridization with the presumably passive band that now crosses the Fermi level at the zone edge and further provides the Fermi surface sheet with the largest area. This drastically deteriorates the nesting of the Fermi surface, and thereby the tendency towards single-stripe AFM order as we discuss below.

In fact, the fully optimized $P4/nmm$ paramagnetic structure obtained in our DFT calculations agrees remarkably well with the experimental one. Specifically, we find the lattice parameters $a = 4.114$ Å and $c = 8.144$ Å with $z_{Si} = 0.108$ and $z_{La} = 0.649$, so that the calculated anion height is $h_{Fe-Si} = 0.88$ Å (i.e. the difference with the experimental lattice parameters is below 0.7 % while the difference with the experimental $h_{Fe-Si}$ is 6 %). This is in striking contrast to the pnictides, in particular LaFeAsO, where such a degree of agreement is only obtained in magnetically ordered solutions —thereby revealing a non-trivial magneto-structural interplay [48–50]. This interplay, however, is absent in LaFeSiO.

To further verify this circumstance, we considered the most relevant magnetic orders and we found in fact a much weaker overall tendency towards magnetism. This is the case even at the generalized gradient approximation (GGA) level, which is known to overestimate the magnetism in the Fe-based superconductors [51]. Specifically, while we find a ferromagnetic solution, this is nearly degenerate with the paramagnetic one and has a very low Fe magnetic moment of $\mu_{Fe} = 0.16\ \mu_B$. Furthermore, the single-stripe AFM solution, characteristic of the pnictides, converged to the paramagnetic ($\mu_{Fe} = 0$) solution. The absence of single-stripe antiferromagnetic solution is indeed totally in tune with the absence of Fermi-surface nesting (Fig. 5c). Still, we find a double-stripe antiferromagnetic solution whose energy difference with respect to the paramagnetic state is just $\Delta E = -5$ meV/Fe with $\mu_{Fe} = 0.58\ \mu_B$ and also a checkerboard one with $\Delta E = -36$ meV/Fe and $\mu_{Fe} = 1.07\ \mu_B$. We note that these magnetization energies are drastically reduced compared to the results obtained assuming LaFeSiO in the reference LaFeAsO structure (i.e. replacing As by Si in LaFeAsO structure) [52]. Consequently, this analysis pinpoints a direct link between the actual unique structure of LaFeSiO and its reduced tendency towards magnetism. Overall, the specific Fermiology and the modest strength of antiferromagnetic spin fluctuations seen in DFT corroborate the conclusions drawn from the NMR results.

## DISCUSSION AND CONCLUSION

In summary, we have reported superconductivity in the new crystallogenide LaFeSiO$_{1-\delta}$. This system displays a drastically reduced anion height $h_{Fe-Si} = 0.94(1)$ Å and yet superconductivity with onset $T_c = 10$ K. In addition, it exhibits relatively weak spin fluctuations, consistent with predictions from first-principles, combined with a non-Fermi-liquid behavior in its normal state. To the best of our knowledge, the conjunction of such structural and superconducting properties is unprecedented in the Fe-based superconducting materials. For this category of unconventional superconductors, there seems to exist a quasi-universal link between structure and $T_c$ that is further connected to the corresponding Fermiology [11, 53]. Thus, the optimal $T_c$ corresponds to having both electron and hole Fermi-surface pockets whose nesting further favors the $s_\pm$-wave mechanism. The pockets, however, may disappear as in the strongly electron doped systems or in the intercalated selenides [53]. In this case, superconductivity is believed to require stronger electronic correlations, eventually leading to a $d$-wave state. LaFeSiO$_{1-\delta}$, however, materializes the opposite situation. Namely, the severe reduction of the anion height is accompanied with a drastic suppression of the initial electron pockets from the Fermi surface. This is obviously detrimental for the $s_\pm$-mechanism, so that the emergence of superconductivity is likely due to stronger correlations, also in tune with its non-Fermi-liquid behavior. However, compared to the chalcogenides, the nature of these correlations is likely different since they originate from a different part of the Fermi surface (i.e. from hole as opposed to electron pockets in the intercalated chalcogenides).

We note that the Fermiology of the initial superconducting crystallogenide LaFeSiH and its fluoride LaFeSiF$_{0.7}$ counterpart still matches that of the reference LaFeAsO material [4, 5, 54]. Namely, even if the out-of-plane dispersion becomes significant in LaFeSiH, the Fermi surface of these crystallogenides display the characteristic electron and hole pockets of the Fe-based superconductors. However, this is not the case in LaFeSiO$_{1-\delta}$ as we described above and a similar situation takes place in LaFeSiF$_{0.1}$ [5]. In both these systems the "canonical" electron pockets undergo a dramatic modification while the effective doping with respect to LaFeSiH is mainly absorbed by the otherwise passive band that gives rise to the heavy 3D hole pocket in LaFeAsO [46]. Consequently, despite their apparent difference in doping, these crystallogenides may well belong to a new superconducting dome in the "Lee plot" where the hole pockets become the essential ingredient as we illustrate in Fig. 6. So, beyond further demonstrating the possibility of Fe-based superconductivity in crystallogenides, our findings challenge the current picture of Fe-based superconductivity and are hence expected to motivate further investigations.

## METHODS

### Synthesis

In order to obtain LaFeSiO$_{1-\delta}$, the LaFeSi precursor was heated either in air, under an oxygen flow or an emulated air flow (Ar 80 %/O$_2$ 20 %) for several days. Different conditions were tried in an attempt to control the oxygen content. However, this has been unsuccessful and essentially the same stoichiometry was obtained in all instances. From a crystallinity point of view, the optimal treatment temperature was found to be 330 °C based on *in-situ* X-ray diffraction (XRD) measurements (Fig. 1 in S.I.). The oxygen uptake was also confirmed qualitatively by thermogravimetric analysis (Fig. 2 in S.I.). The purest LaFeSiO$_{1-\delta}$ sample batch had a phase purity of 96(1) % and was obtained specifically by heating the precursor for 3 days at 330 °C in an emulated air flow consisting of 80 % Ar and 20 % O$_2$. The ramp which was used for both heating and cooling was 10 °C/min. Secondary phases, already contained in the LaFeSi precursor, persist through the oxygenation process. Namely, the ferromagnetic La(Fe,Si)$_{13}$ and the paramagnetic LaFe$_2$Si$_2$ and correspond to ~2.5(5)% and ~1.5(5)% of the oxygenated sample respectively. The phase purity was estimated by performing a Rietveld fit of X-ray diffraction data (Fig. 3 in S.I.).

### Resistivity

The resistivity measurements shown in Fig. 3 (and Fig. 10 in S.I.) correspond to our 96 % pure LaFeSiO$_{1-\delta}$ batch. Resistivity was measured on a sample grain of approximately 150 $\mu$m x 50 $\mu$m x 50 $\mu$m. The grain was measured using a 4 circle diffractometer ($\lambda$(K$_\alpha$(Mo) = 0.71 Å) revealing it to be single phase, consisting of a hand full of 1111-type grains. The azimutally integrated data can be indexed with the LaFeSiO$_{1-\delta}$ phase determined by NPD, linking the structure and superconducting properties.

### Magnetization

The magnetization was measured using the same sample batch as for the resistivity in a Quantum Design MPMS-XL. The sample holder was a thin straw wherein a small pellet of 25.9 mg was fixed using plastic film. The sample was centered without applying an external field. It was then brought to 2 K where upon the field sweep was carried out. The sample was then heated to 300 K in no applied field and cooled to 15 K before once again measuring M(H).

### Neutron powder diffraction

The crystal structure was investigated using neutron powder diffraction on the D1B instrument [55] at the ILL using a wavelength of $\lambda$ = 1.28 Å. For this experiment, we used a large sample containing 67(2) % of LaFeSiO$_{1-\delta}$, ~ 29(1) % of unreacted LaFeSi and ~ 4.0(5) % of LaFe$_2$Si$_2$. The crystal structure of LaFeSiO$_{1-\delta}$ and the proportions of the phases were refined using the Rietveld method in the FULLPROF software [56].

### X-ray powder diffraction

All samples which were produced were investigated by powder XRD, using a D8 Endeavor diffractometer with a K$_{\alpha1,\alpha2}$(Cu) source. All samples showed similar unit cell parameters.

### Electron diffraction and energy-dispersive spectroscopy

The TEM analysis was performed on a specimen prepared by suspending a small amount of powder in ethanol, and depositing a drop of the liquid on a copper grid, covered by a holey carbon membrane. The microscope used was a Philips CM300ST (LaB$_6$, 300 kV) equipped with a F416 TVIPS CMOS camera and a Bruker Silicon Drift Energy Dispersive X-ray Spectroscopy (EDX) detector. The 3D electron diffraction (ED) study was performed with a tomography sample holder allowing a tilt range of ± 50 °, using the method described by S. Kodjikian and H. Klein [57]. ED dataset processing was performed using PETS program, and the crystal structure model was calculated by the charge flipping algorithm [58] with the Superflip program [59] in the computing system JANA2006 [60].

### Nuclear magnetic resonance

$^{29}$Si measurements were performed in a fixed field of 15 T from a superconducting coil, using a home-built heterodyne spectrometer. The field value was calibrated using metallic Cu from the NMR pick-up coil. Knight shift values are given with respect to the bare $^{29}$Si resonance. Spectra were obtained by adding appropriately-spaced Fourier transforms of the spin-echo signal. The spin-lattice relaxation time $T_1$ was measured by the saturation-recovery method and the recoveries were fit to the theoretical law for magnetic relaxation of a nuclear spin 1/2: $M(t) = M(\infty)(1 - \exp(-(t/T_1)^\beta))$, modified by an ad-hoc stretching exponent $\beta$ in order to account for a distribution of $T_1$ values [61].

### Electronic structure calculations

The main calculations were performed using the all-electron code WIEN2k [62] based on the full-potential augmented plane-wave plus local orbitals method (APW+LO). We considered the Perdew-Burke-Ernzerhof (PBE) form of the generalized gradient approximation (GGA) [63] and used muffin-tin radii of (La) 2.30 a.u., (Fe) 2.10 a.u., (Si) 2.10 a.u., and (O) 1.80 a.u. with a plane-wave cutoff $R_{MT}K_{max}$ = 7.0. Additional calculations were performed with Quantum Espresso [64] using the norm-conserving ONCVPSP pseudopotentials from Dojo [65, 66].

## DATA AVAILABILITY

NPD data used for Fig. 2.b and Table I are available at ref.[55]. The other data that support the findings of this study are available from the corresponding author upon reasonable request.

## ACKNOWLEDGEMENTS

We want to thank Jacques Pecaut for his measurement on a 4-circle diffractometer of the sample grain measured in resistivity. We thank Rémy Bruyere, Paul Chometon and Frédéric Gay for assistance in high-temperature XRD, thermogravimetric analysis and resistivity measurements, respectively. We thank the ILL for providing beamtime at the D1B instrument.

This work was supported by the ANR-18-CE30- 0018-03 Ironman grant and the French state funds ANR-10-LABX-51-01 (Labex LANEF du Programme d'Investissements d'Avenir). F.B. acknowledges support from Cineca ISCRA-C project "IsC78-NICKSUP-HP10C91RDL". A.C. acknowledges support from the Visiting Professor/Scientist 2019 program founded by the Regione Autonoma Sardegna.

## AUTHOR CONTRIBUTIONS

P.T., S.T. and A.C. designed the research; M.F.H., J.-B.V., C.L., H.M., M.H.J., F.B., E.G, V.N. M.-A.M. and A.C. performed research; M.F.H., J.-B.V., C.L., E.G., A.S. and M.H.J. analysed data, M.F.H, M.H.J., A.C. and P.T wrote the paper.

## COMPETING INTERESTS

The authors declare no competing interests.

## ADDITIONAL INFORMATION

**Correspondence** and requests for materials should be addressed to P.T.

| $P4/nmm$ (#129, origin 2) | | | | |
|---|---|---|---|---|
| T = 300 K, $a$ = 4.1085(4) Å, $c$ = 8.132(2) Å. | | | | |
| Atom | Wyckoff pos. | $x$ $y$ | $z$ | Occ. |
| La | $2c$ | 1/4 1/4 | 0.6526(9) | 1 |
| Fe | $2a$ | 3/4 1/4 | 0 | 1 |
| Si | $2c$ | 1/4 1/4 | 0.116(2) | 1 |
| O | $2b$ | 3/4 1/4 | 1/2 | 0.90(2) |

TABLE I. LaFeSiO$_{1-\delta}$ refined crystal structure at 300 K from NPD data (Bragg R-factor = 5.05).

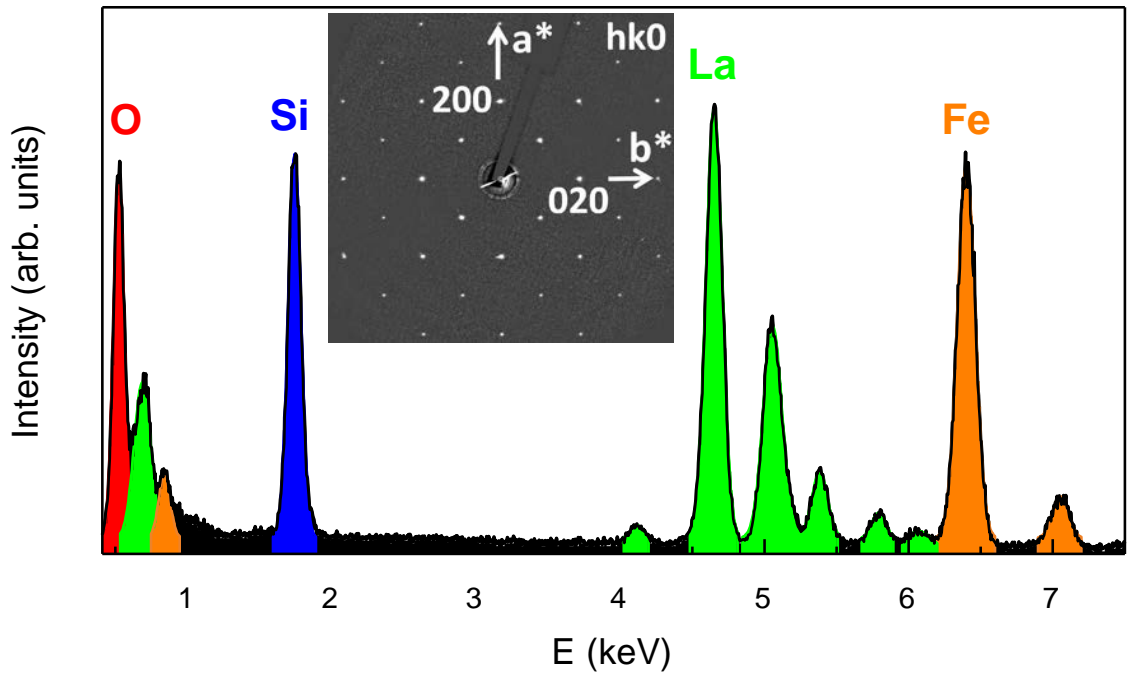

FIG. 1. EDX spectrum recorded on one crystallite showing an oxygen peak at 0.525 keV in the LaFeSi matrix. **Inset:** hk0-cut of the reciprocal space indexed in P4/nmm space group of LaFeSiO$_{1-\delta}$.

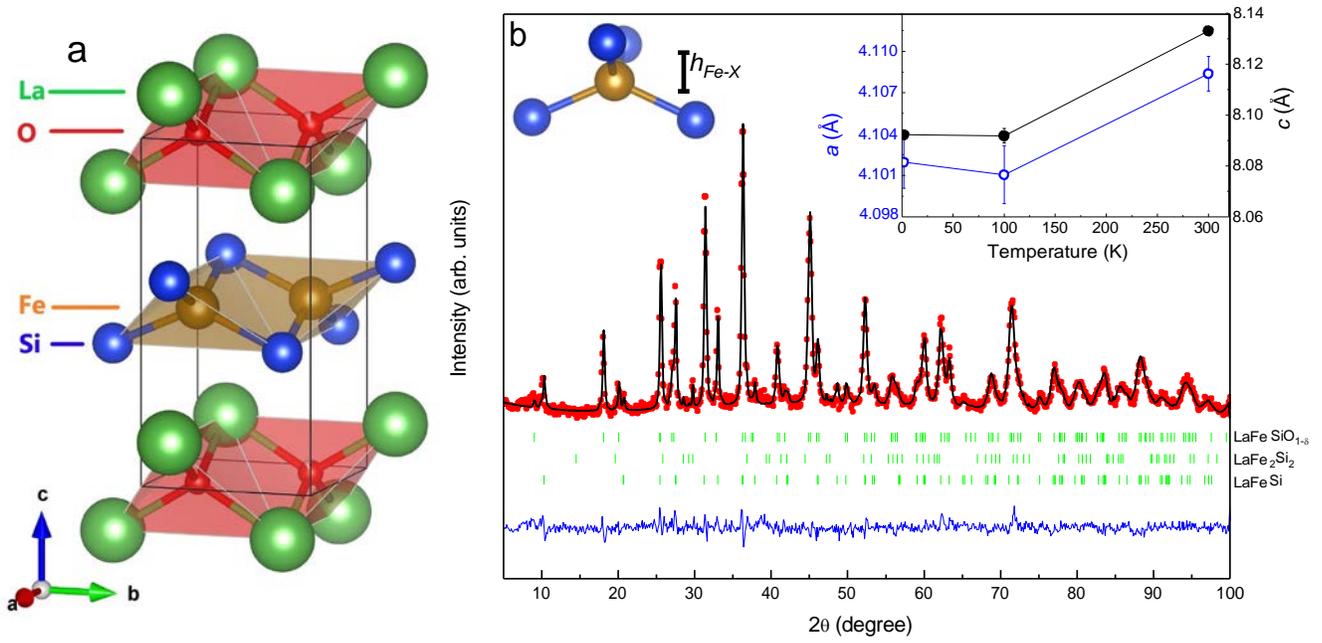

FIG. 2. **a:** The crystal structure as determined by NPD. The structure is shifted by (0,0,0.5) to emphasize the Fe containing layer. **b:** Rietveld refinement of the NPD data collected at 300 K at the D1B instrument of ILL. Three phases are included: LaFeSiO$_{1-\delta}$ , LaFe$_2$Si$_2$ and unreacted LaFeSi (from top to bottom). **Inset:** Temperature dependence of $a$ and $c$ lattice parameters.

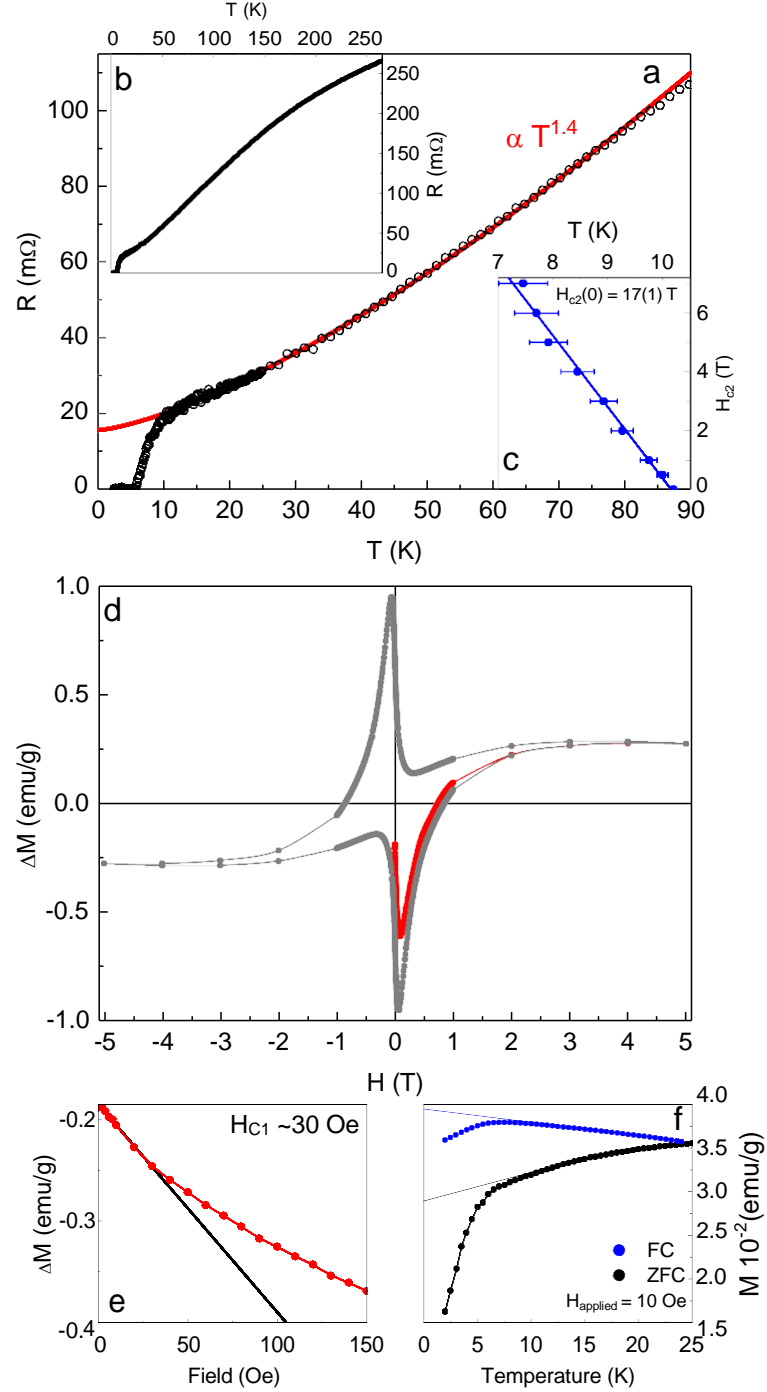

FIG. 3. **a:** Electrical resistance of LaFeSiO$_{1-\delta}$ as a function of temperature showing superconductivity with onset $T_c$ = 10 K and non-Fermi liquid $T^{1.4}$ behavior in the normal state. **b:** Extended plot of the resistance up to room temperature. **c:** Superconducting transition temperature versus applied field. The fit to the WHH formula gives $H_{c2}(0)$ = 17 T. **d:** Superconducting hysteresis loop obtained by difference from the magnetization measured at 2 K and 15 K for a cold pressed cylinder of polycrystalline LaFeSiO$_{1-\delta}$. **e:** A zoom of the initial part of **d**, fitted with a linear expression to obtain the susceptibility. **f:** The field-cooled/zero-field-cooled magnetization curves measured on the same pellet of LaFeSiO$_{1-\delta}$.

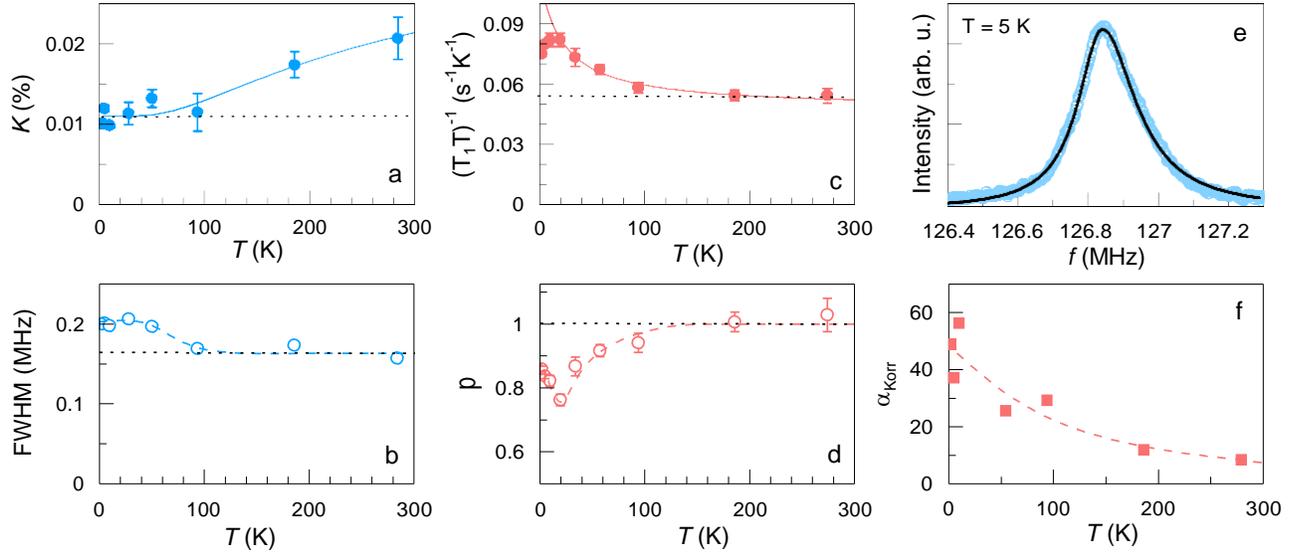

FIG. 4. **a:** Knight shift defined from the peak position of the $^{29}$Si resonance shown in panel **e:**. The continuous line represents the activated form $K = A + B \exp(-\Delta/k_B T)$, as observed in various Fe-based superconductors, with $\Delta = 250$ K here. **b:** Full width at half maximum of the $^{29}$Si resonance shown in panel **e:**. The line broadening arises from a distribution of Knight shifts values. The dashed line is a guide to the eye. **c:** Spin-lattice relaxation rate divided by temperature. The continuous line is a fit to the Curie-Weiss form $a + c/(T + \theta)$ with $\theta = 22$ K. **d:** Stretching exponent $\beta$ used to fit the nuclear recoveries in a $T_1$ experiment (see Methods). $\beta$ provides a measure of the width of the distribution of $T_1$ values. **e:** $^{29}$Si NMR line at $T = 5$ K in a field of 15 T. The continuous black line is a fit to an asymmetric Lorentzian form. The slight asymmetry may arise from Knight shift anisotropy as all directions contribute to the spectrum in this powder sample. **f:** Ratio $a_{\text{Korr}} = \hbar \gamma^2_{\;\;B\;n}(1/(T_1 T K^2)$. $a_{\text{Korr}} \gg 1$ is evidence of predominant antiferromagnetic fluctuations. The dashed lines in panels b, d and f are guides to the eye and the dotted lines in all the panels represent $T$ independent behavior.

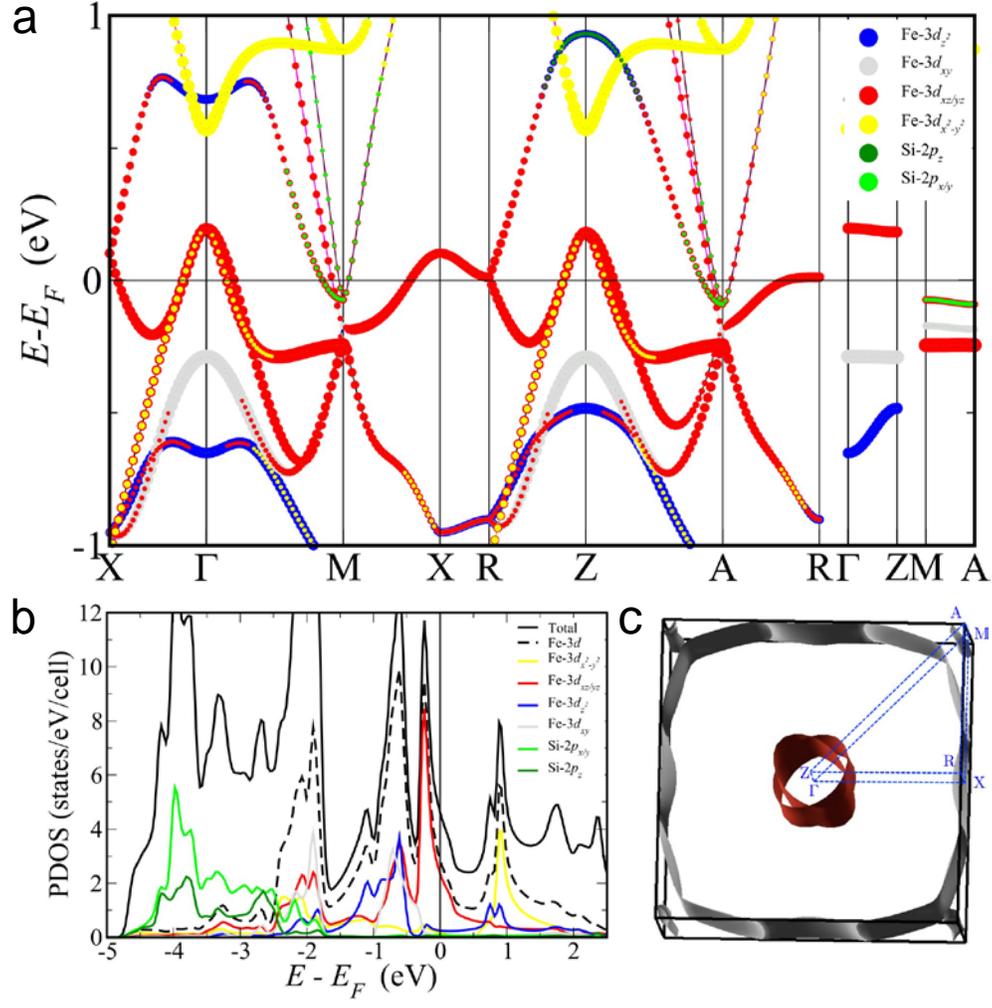

FIG. 5. Calculated electronic structure of LaFeSiO. **a:** Orbital-resolved band structure along the high-symmetry directions of the P4/$nmm$ Brillouin zone. **b:** Orbital-resolved density of states. **c:** Perspective view of Fermi surface computed on the basis of the experimental structure in Table I. The labels indicate the high-symmetry points and lines correspond to the $k$-path in **a** .

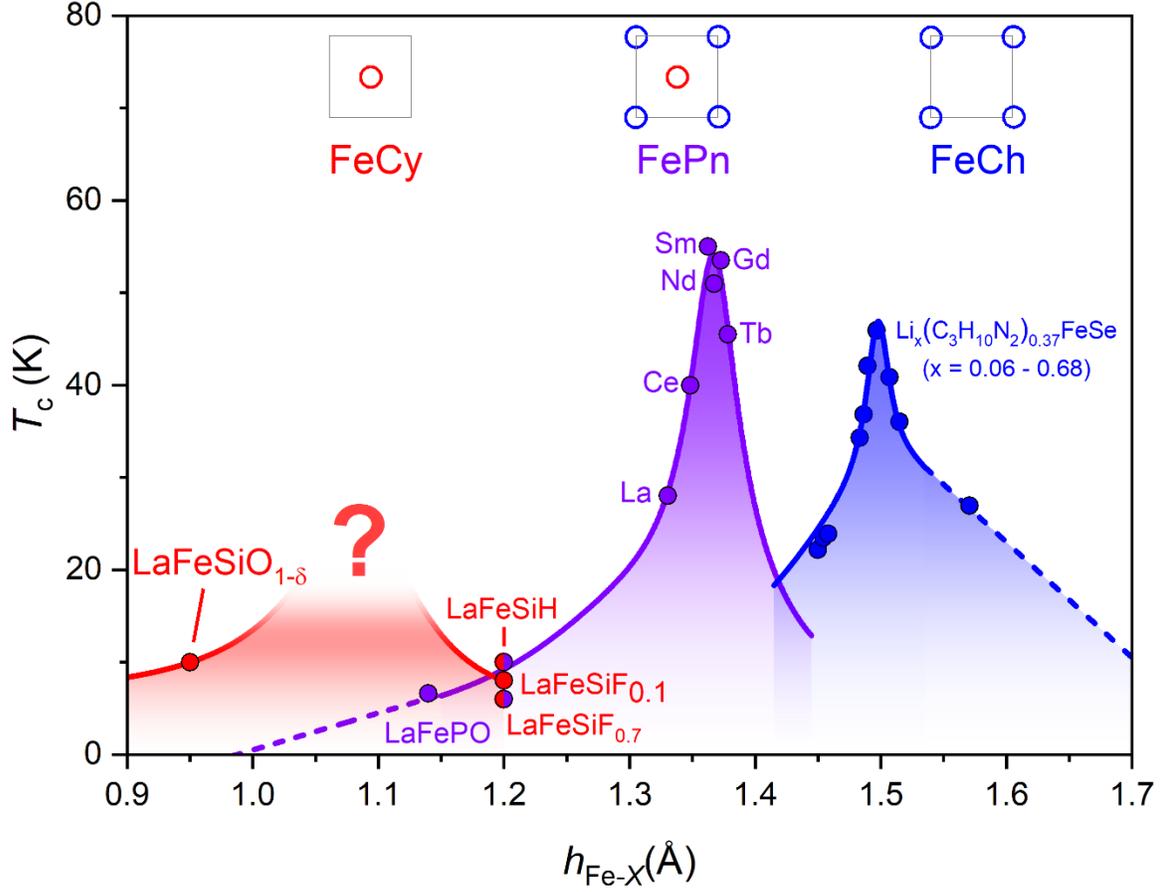

FIG. 6. Superconducting transition temperature $T_c$ as a function of the anion height from the Fe plane $h_{Fe-X}$ for pnictides (Pn) in purple (labelled with RE for REFeAsO$_{1-x}$(H,F)$_x$) [67] (and references therein), heavily electron doped chalcogenides (Ch) in blue [53] and crystallogenides (Cy) in red (LaFeSiO$_{1-\delta}$ (this work), LaFeSiF$_x$ [5] and LaFeSiH [4]). The sketches illustrate the simplified Fermi surfaces of these materials and LaFeSiH and LaFeSiF$_{0.7}$ are marked in red/purple to indicate that the Fermiology of these crystallogneides bear resemblance to the pnictides. The $T_c$ of the pnictides and chalcogenides peaks at different $h_{Fe-X}$ values, which can be ascribed to their different Fermiology (and hence pairing mechanism). The new superconducting crystallogenide LaFeSiO$_{1-\delta}$ reported in this work provides yet another Fermiology and appears above the tail of the pnictide $T_c(h_{Fe-X})$ curve. This might reveal another FeSi-based superconducting dome.